\documentclass[twocolumn,superscriptaddress,amsmath,amssymb]{revtex4}
\pdfoutput=1
\usepackage{graphicx}
\usepackage{dcolumn,xcolor}
\usepackage{amsmath} 
\usepackage{amssymb}
\usepackage{amsfonts}
\usepackage{bm}
\usepackage{overpic}

\begin{document}

\title{Metabolic cascades of energy}

\author{M.A.~Fardin}
\altaffiliation[Corresponding author ]{}
\email{marc-antoine.fardin@ijm.fr}
\affiliation{Universit\'{e} de Paris, CNRS, Institut Jacques Monod, F-75006 Paris, France}
\affiliation{The Academy of Bradylogists.}

\date{\today}

\begin{abstract}
Life has a special status, it even has its own science: biology. In many ways, the logic of life seems to differ from that of atoms, molecules, planets, or any other `inanimate object'. However, life is increasingly measured using quantities shared by all sciences, like mass, force, energy or power. An analysis of the dimensions of these quantities provides powerful ways to infer the relationships they might have with one another. Here we show that a dimensional analysis of the metabolic laws connecting the characteristic powers and masses of living organisms offers new ways to understand the deep connections between the chemistry of microscopic molecules and the physics of macroscopic objects bound by gravity. This analysis reveals a link between metabolism and the cascades of energy observed in turbulent flows, opening new perspectives for both fields. 
\end{abstract}

\maketitle

The rate of energy expenditure of living organisms is usually called the `metabolic rate'~\cite{Glazier2010}. As an energy per unit time it is a power $P$, with dimensions $\mathcal{M}.\mathcal{L}^2.\mathcal{T}^{-3}$, and its standard unit is the Watt. It can be expressed in kg.m$^2$.s$^{-3}$, but is more often expressed in J/s, or kcal/day, i.e. in some unit of energy over some unit of time. Beyond any particular choice of units, the dimensions remain. 

There has been a considerable amount of discussion in the literature on the relationship between the metabolic rate $P$ and the mass $m$ of the organisms across kingdoms and scales~\cite{Brown2004,West2005,Glazier2010,Makarieva2008,Hatton2019}.  Obviously, an elephant eats more than a mouse, but how much exactly? Such metabolic laws are particularly interesting because they can be used to infer other relationships between mass and growth rate, mortality rate, or abundance, to cite just a few examples~\cite{Brown2004,Hatton2019}. 

There is a mass in the dimensions of any power ($\mathcal{M}.\mathcal{L}^2.\mathcal{T}^{-3}$), and so it is natural to ask how this dimension of mass may be connected to the mass $m$. Among the many types of relationships between $P$ and $m$ that have been investigated, one continues to have a particularly stimulating impact: $P\sim m^{3/4}$. This relationship is known as Kleiber's law~\citep{Kleiber1932}. In some instances, it has been verified to a large extent~\cite{West2005}, in others it has served as a point of comparison to design alternative metabolic relationships~\cite{Glazier2005}. Theoretical monuments have been built in order to explain or refute it. Our purpose here is not to review this extensive literature, but to show that a few dimensional arguments can be used to navigate more easily through it. Once carried out, dimensional analysis actually reveals stimulating connections between metabolic laws and turbulence spectra. Consequences of these connections run wide, as we shall progressively show throughout this article. 

\section*{The metabolic law behind turbulence}
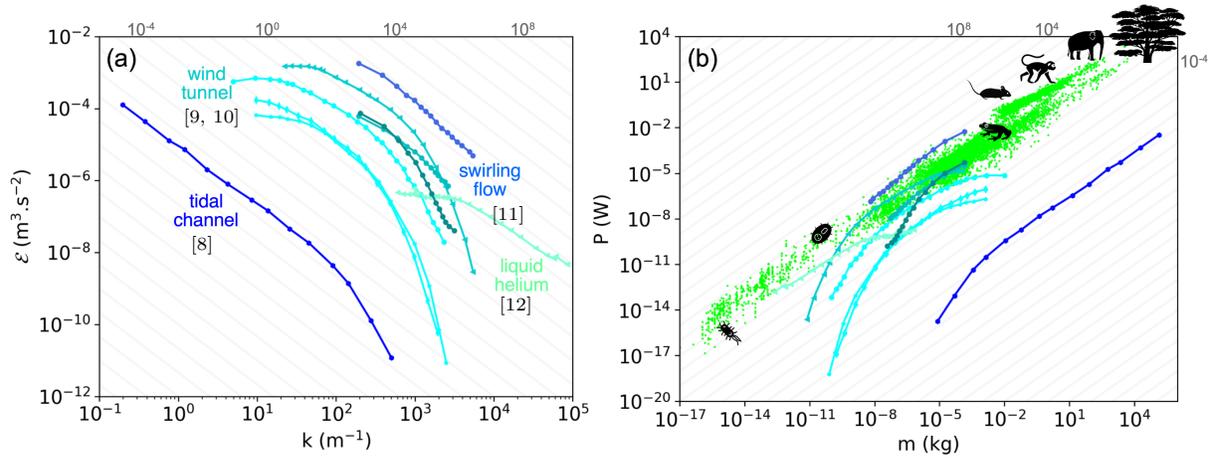
\begin{figure*}
\centering
\begin{overpic}[abs,unit=1mm,width=0.9\linewidth]{ 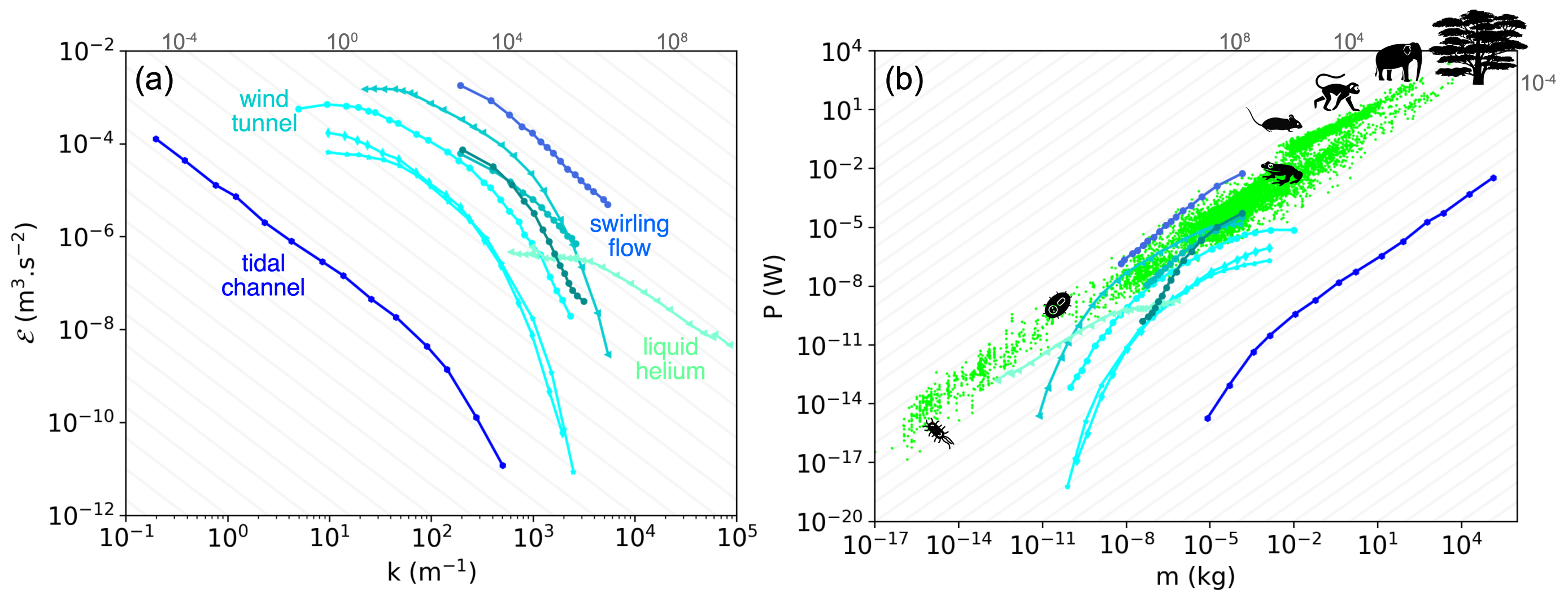}
\put(25,28){\scriptsize\cite{Grant1962}}
\put(24,45){\scriptsize\cite{Uberoi1963,Comte1971}}
\put(65,32){\scriptsize\cite{Debue2018}}
\put(66,20){\scriptsize\cite{Salort2010}}
\end{overpic}
\caption{Conversion from power spectrum (a) to metabolic law (b) for representative data from the literature on the turbulence of various fluids in different flow geometries. (a) The wind tunnel data correspond to $\rho\simeq 1.2$~kg/m$^3$~\cite{Uberoi1963,Comte1971}. The turbulence was excited by a grid, and the different trends corresponds to different types of grids or to measurement realized at different distances downstream from the grid. The turbulence data of liquid helium correspond to $\rho\simeq 148$~kg/m$^3$~\cite{Salort2010}. The tidal channel data correspond to underwater turbulence in the sea, with $\rho\simeq 10^3$~kg/m$^3$~\cite{Grant1962}. The swirling flow data were obtained on mixtures of water and glycerol, with $\rho\simeq 998$, 1000, 1166~kg/m$^3$~\cite{Debue2018}. (b) The same turbulence data are plotted as metabolic laws, together with the metabolic rates of a vast range of organisms from bacteria to large mammals and trees, reproduced from Hatton \textit{et al.}~\cite{Hatton2019}. More details on the different taxonomic groups will be given in Fig.~\ref{fig2}. On both graphs (a) and (b), the diagonal grey lines correspond to different values of transfer rate $\epsilon$, expressed in m$^2$.s$^{-3}$.}
\label{fig1}
\end{figure*}
Let us assume that the relationship between metabolic rate $P$ and mass $m$ is a power law. In doing so, we neglect the `curvature' that data might present in logarithmic scale~\citep{Kolokotrones2010}. This amounts to neglecting so-called `logarithmic corrections', providing a degree of approximation that we acknowledge, but that will be sufficient for our purpose. We will write $P\sim m^{\alpha}$, where the sign `$\sim$' means that if one were to plot log(P) vs log(m) one would find a straight line of slope $\alpha$. The intercept log($K$) with the log(P)-axis would give rise to the prefactor of the power-law, such that we have $P=K m^{\alpha}$. We will call such $P(m)$ relationships `metabolic laws', even when we venture outside the realm of living organisms. 

Whereas a lot of contributors on this topic have devoted their efforts to finding the value of $\alpha$, the prefactor $K$ has not received the attention it deserves. Both $K$ and $\alpha$ are intimately connected if one analyses the dimensions of the equation. As is customary, we will write $[K]$ to denote the dimensions of a quantity $K$. Since $[P]=\mathcal{M}.\mathcal{L}^2.\mathcal{T}^{-3}$, and $[m]=\mathcal{M}$, then $[K]=\mathcal{M}^{1-\alpha}.\mathcal{L}^2.\mathcal{T}^{-3}$. This equation is true regardless of the value of $\alpha$. 

One value of $\alpha$ is special by construction. Indeed, $\alpha=1$ removes the mass dimension of $K$, $[K]=\mathcal{M}^0.\mathcal{L}^2.\mathcal{T}^{-3}$. This would give a metabolic rate proportional to mass:
\begin{equation}
P=K_1 m \label{G0}
\end{equation}
\noindent where we use the subscript 1 to keep in mind that this coefficient corresponds to $\alpha=1$. In this elementary case, what could be the interpretation of the proportionality constant $K_1$? Because the dimensions of $K_1$ only involve space and time we say that it is a kinematic quantity. For comparison, a speed would have dimensions $\mathcal{L}.\mathcal{T}^{-1}$, a coefficient of diffusion would have dimensions $\mathcal{L}^2.\mathcal{T}^{-1}$, or an acceleration would have dimensions $\mathcal{L}.\mathcal{T}^{-2}$. In contrast, $K_1$ has dimensions $\mathcal{L}^2.\mathcal{T}^{-3}$. This quantity is not as well known as a speed or a coefficient of diffusion, but it appears prominently in the field of turbulence~\cite{Kolmogorov1941,Frisch1995}. In this context, one studies the complex flows made up of many swirls as seen for instance in some tumultuous rivers or in puffs of smokes and clouds. 

Almost tautologically, one can call $K_1$ the power per unit mass, since $K_1=P/m$. In the context of turbulence, a coefficient with the same dimensions than $K_1$ is usually called the `rate of energy transfer' or the `dissipation rate', and the symbol used is $\epsilon$. This rate of dissipation is the parameter of the so-called `Kolmogorov energy spectrum', which is usually expressed as: 
\begin{equation}
\mathcal{E}(k) = C_K \epsilon^\frac{2}{3} k^{-\frac{5}{3}} \label{Kolmo}
\end{equation}
This equation might seem odd to those unfamiliar with it, but we shall see that it is essentially equivalent to $P=\epsilon m$. The term `$C_K$' is the dimensionless `Kolmogorov constant', with a value around $\frac{1}{2}$~\cite{Sreenivasan1995}. In the following, we will ignore such dimensionless coefficient `of order 1' and write $\mathcal{E}(k) \propto \epsilon^\frac{2}{3} k^{-\frac{5}{3}}$, where the sign `$\propto$' is here to remind us that we neglect these coefficients. The symbol `$k$' refers to the `wavenumber', with dimensions $[k]=\mathcal{L}^{-1}$. The `energy spectrum' $\mathcal{E}(k)$ is a kinematic quantity constructed from the fluctuations of the velocity of the flow over a size $\ell\propto k^{-1}$. Its dimensions are $[\mathcal{E}]=\mathcal{L}^3.\mathcal{T}^{-2}$. In the context of turbulence, one typically measures how the many-sized vortices in the fluid affect its flow. The apparent complexity of the scaling $\mathcal{E}(k)\sim k^{-\frac{5}{3}}$ comes from translating simple assumptions to the particular frame of reference most commonly used by experiments. 

To show how Kolmogorov's energy spectrum can be expressed as a metabolic law, we can first translate the variable $k$ into a mass $m$. If the spectrum depends on $k$, then it depends on the size $\ell\propto k^{-1}$, and on the mass $m\propto \rho \ell^3$ of a parcel of fluid of size $\ell$  and density $\rho$. Thus, we have $k \propto (\rho/m)^\frac{1}{3}$. On the left-hand side of Eq.~\ref{Kolmo}, how can we connect $\mathcal{E}(k)$ to $P$, $m$ and $\rho$? The answer is given by dimensional analysis: 
\begin{equation}
\mathcal{E} \propto m^\beta P^\gamma \rho^\delta
\label{startE}
\end{equation}
An equation like this is just a formal way of saying that $\mathcal{E}$ depends on the mass $m$, the power $P$, and the density $\rho$. We can translate the equation into an equation on the dimensions: 
\begin{align}
\mathcal{L}^3.\mathcal{T}^{-2} &= \mathcal{M}^\beta (\mathcal{M}.\mathcal{L}^{2}.\mathcal{T}^{-3})^\gamma (\mathcal{M}.\mathcal{L}^{-3})^\delta \nonumber\\
\mathcal{M}^0.\mathcal{L}^3.\mathcal{T}^{-2} &= \mathcal{M}^{\beta + \gamma +\delta}.\mathcal{L}^{2\gamma-3\delta}.\mathcal{T}^{-3\gamma}
\end{align}
We then have a system of three equations, one for each dimension: 
\begin{equation}
\left\{
  \begin{array}{lr}
    \mathcal{M}:& \beta + \gamma +\delta =0\\
    \mathcal{L}:&  2\gamma-3\delta= 3\\
    \mathcal{T}:&  -3\gamma= -2
  \end{array}
\right.
\end{equation}
This system can easily be solved to get $\beta=-\frac{1}{9}$, $\gamma=\frac{2}{3}$ and $\delta=-\frac{5}{9}$. We thus have translation formulas to go from turbulence spectra to metabolic laws and vice-versa: 
\begin{equation}
  \begin{array}{ll}
    k &\equiv (\rho/m)^\frac{1}{3} \\
	\mathcal{E} &\equiv m^{-\frac{1}{9}} P^\frac{2}{3} \rho^{-\frac{5}{9}}
  \end{array}  \Leftrightarrow
  \begin{array}{ll}
    m &\equiv \rho k^{-3} \\
	P &\equiv \rho \mathcal{E}^{\frac{3}{2}} k^{-\frac{1}{2}}
  \end{array}
  \label{translate}
\end{equation}
By translating the energy spectrum and wavenumber in Eq.~\ref{Kolmo} one recovers Eq.~\ref{G0}, with the symbol $\epsilon$ used instead of $K_1$:
\begin{equation}
\mathcal{E} \propto  \epsilon^\frac{2}{3} k^{-\frac{5}{3}}
\Leftrightarrow m^{-\frac{1}{9}} P^\frac{2}{3} \rho^{-\frac{5}{9}}\propto \epsilon^\frac{2}{3} (\rho/m)^{-\frac{5}{9}}
\Leftrightarrow P\propto \epsilon m
\end{equation}

Both equations encompass the same statement but expressed in two distinct systems of units. The metabolic variables $P$ and $m$ provide a quite synthetic way to express Kolmogorov's spectrum. 

In practice, the equivalence between the metabolic law in Eq.~\ref{G0} and the energy spectrum in Eq.~\ref{Kolmo} means that for a given density $\rho$ we can use the translation formulas to express a metabolic law into an energy spectrum, or a spectrum into a metabolic law. These are just two equivalent ways to represent data. For instance, in Fig.~\ref{fig1} we compare a few turbulence spectra from the literature expressed as $\mathcal{E}$ vs $k$, or as $P$ vs $m$. 

As is apparent, the data do not systematically follow the $\mathcal{E}\sim k^{-\frac{5}{3}}$ scaling, i.e. the data points extend beyond $P\sim m$. Generally, the data points follow $P\sim m$ only in the range $m_1 <m<m_2$, i.e. between $\ell_1 <k^{-1}<\ell_2$, with $m_1\propto \rho \ell_1^3$ and $m_2\propto\rho \ell_2^3$. The values of these bounds vary from one experiment to another, and so does the value of $\epsilon$ characterizing the data between these bounds. For instance, in Fig.~\ref{fig1}, the data from the tidal channel correspond to $\epsilon\simeq 5~10^{-8}$~m$^2$.s$^{-3}$ for $\ell_1 <k^{-1}$, with $\ell_1 \simeq 1$~cm~\cite{Grant1962}. No bound $\ell_2$ could be identified. In contrast, some of the wind tunnel data provide both bounds. For instance, the cyan dots correspond to $\epsilon\simeq 10^{-1}$~m$^2$.s$^{-3}$ for $\ell_1 <k^{-1}<\ell_2$, with $\ell_1 \simeq 2$~mm and $\ell_2 \simeq 2$~cm~\cite{Comte1971}. For a particular experiment, the bounds $\ell_1$ and $\ell_2$ respectively correspond to the crossovers toward what are usually called the `dissipation scale' and `the integral scale'. We will more neutrally call these the `microscopic' and `macroscopic' scales. 

What sets the values of the transfer rate $\epsilon$ and of the bounds $\ell_1$ and $\ell_2$ in a particular experiment? The sizes $\ell_1$ and $\ell_2$ are identified graphically as the points of departure from the scaling $\mathcal{E}\sim k^{-\frac{5}{3}}$. They are the last points abiding to the scaling, and so we necessarily have: 
\begin{equation}
\epsilon \propto \Big( \frac{\mathcal{E}_i^3}{\ell_i^5} \Big)^\frac{1}{2} \propto \frac{P_i}{m_i}
\label{bounds}
\end{equation}
\noindent where $i=1,2$. These expressions explain why $\epsilon$ is often called `the dissipation rate'. Indeed, if $\epsilon$ is expressed from the lower bound $\ell_1$ then it is associated with the power $P_1$, which is indeed where dissipation occurs in standard fluids. However, one could very well call $\epsilon$ the `injection rate' if defined from the integral scale where a power $P_2$ is generated. In practice, the control parameters for a given experiment are usually connected to $P_2$ and $\ell_2$.

\section*{Universality of the metabolic power density}
In turbulent studies, the value of $\epsilon$ typically varies from one experiment to another depending both on the fluid properties and on the way energy is fueled into the system at large scales. It can be expressed externally from the macroscopic scale, or internally from the microscopic scale. This versatility is no surprise if one writes $\epsilon\propto P/m \propto P/\rho\ell^3$. We can define the power density $\Pi \propto P/\ell^3$, i.e. the power per unit volume, in the same way that $\rho\propto m/\ell^3$ is the mass per unit volume. With these definitions, we have: 
\begin{equation}
\epsilon \propto \frac{\Pi}{\rho}
\end{equation}
Whereas $P$ and $m$ are extensive properties, $\Pi$ and $\rho$ are intensive, i.e. they are expected to remain constant over a broad range of sizes. For a fluid like water in normal conditions, the density will remain constant over a very large range of size, down to atomic scale. However, the power density $\Pi$ will be constant over a smaller range, set by $\ell_1$ and $\ell_2$. In stark contrast, living organisms seem to broadly share the same universal metabolic power density, over a very large range of sizes~\cite{Makarieva2008,Hatton2019}. 

In Fig.~\ref{fig1}b we plot the metabolic rates of a vast range of living organisms obtained from Hatton \textit{et al.}~\cite{Hatton2019}. On the same $P$ vs $m$ graph, we also reproduce some turbulence data from Fig.~\ref{fig1}a. The metabolic data as well as the turbulence data express how the rate of energy transfer (i.e. the power $P$) depends on the mass $m$ (and so size $\ell\propto (m/\rho)^\frac{1}{3}$). In this analogy, each organism is like a fluid vortex, or each vortex is like an organism.   

For living organisms, the rate of energy transfer per unit mass $\epsilon$ typically varies from 0.1 to 10~W/kg~\cite{Hatton2019}, a small range in comparison to turbulence studies. If we consider a rough average $\epsilon_0\simeq 1$~W/kg, and a density always close to that of water $\rho_0\simeq 10^3$~kg/m$^3$, we obtain a fairly universal metabolic power density $\Pi_0\simeq 10^3$~W/m$^3$. How can this value be understood? The insight from turbulence studies given in Eq.~\ref{bounds} would suggest that one can obtain equivalent expressions for $\Pi_0$ (and so $\epsilon_0$) by connecting it to the bounds of the range of validity of the scaling $P\sim m$. A microscopic expression of $\Pi_0$ can be obtained by considering the smallest living organisms, whereas a macroscopic expression can be obtained by considering the largest ones. This is what we shall do now.   

\section*{Microscopic scale and viscosity}
Using the power density, Kolmogorov's scaling can be written as $\mathcal{E}\propto (\Pi/\rho)^\frac{2}{3} k^{-\frac{5}{3}}$. Using the power and mass variables, this is simply $P\propto(\Pi/\rho)m $, i.e. it is just expressing the extensive-to-intensive relationship $P/m \propto \Pi/\rho$. Part of Kolmogorov's insight was to supplement the intensive quantities $\Pi$ and $\rho$ by a third quantity $\eta$ ($\mathcal{M}.\mathcal{L}^{-1}.\mathcal{T}^{-1}$), which is the viscosity of the fluid~\cite{Kolmogorov1941}. With this last quantity, one can build a complete set of units, equivalent to $\mathcal{M}$, $\mathcal{L}$ and $\mathcal{T}$. Using dimensional analysis one can show that $\Pi$, $\rho$ and $\eta$ can be combined in unique ways to construct a time scale $\tau_K$, a length $\ell_K$ and a mass $m_K \propto \rho\ell_K^3$: 
\begin{align}
\ell_K &\propto \Big(\frac{\eta^3}{\rho^2 \Pi}\Big)^\frac{1}{4}\\
\tau_K &\propto \Big(\frac{\eta}{\Pi}\Big)^\frac{1}{2}
\end{align}
Note that one can recast the first equation as $\ell_K \propto (\nu^3 /\epsilon)^\frac{1}{4}$~\cite{Frisch1995}, where $\nu\propto \eta/\rho$ is the kinematic viscosity (momentum diffusivity). One can also express Kolmogorov's length as $\ell_K \propto (\nu \tau_K)^\frac{1}{2}$ or $\ell_K \propto (\epsilon/\tau_K)^\frac{1}{2}$. For the turbulence data in Fig.~\ref{fig1}, the Kolmogorov's lengths ranged from a few microns to a few millimeters~\cite{Grant1962,Uberoi1963,Comte1971,Debue2018}. 

For the metabolic scaling of organisms, we can use the reference values $\Pi_0$ and $\rho_0$ to infer $\ell_K\simeq 5~10^{-3} \eta^\frac{3}{4}$. In other words, the Kolmogorov length associated with the average parameters of the metabolic scaling is directly connected to the value of the viscosity $\eta$. The smallest bacteria in Fig.~\ref{fig1}b correspond to a mass $m\simeq 10^{-17}$~kg and size $\ell\simeq 10^{-7}$~m. The average metabolic law $P\sim m$ extends at least down to this scale, so it is the highest possible value for  Kolmogorov's size, which would then correspond to a viscosity $\eta\simeq 5~10^{-7}$~Pa.s. Lower choices of $\ell_K$, would yield even smaller values of viscosity. For comparison, the viscosity of water is two thousand times larger than the inferred viscosity from the size of bacteria. Blindly applying Kolmogorov's units leads to aberrant values of viscosity, far from the material properties of living organisms. We need a different interpretation of the microscopic scale. 

\section*{Microscopic scale and thermal energy}
\begin{figure*}
\centering
\begin{overpic}[abs,unit=1mm,width=0.9\linewidth]{ 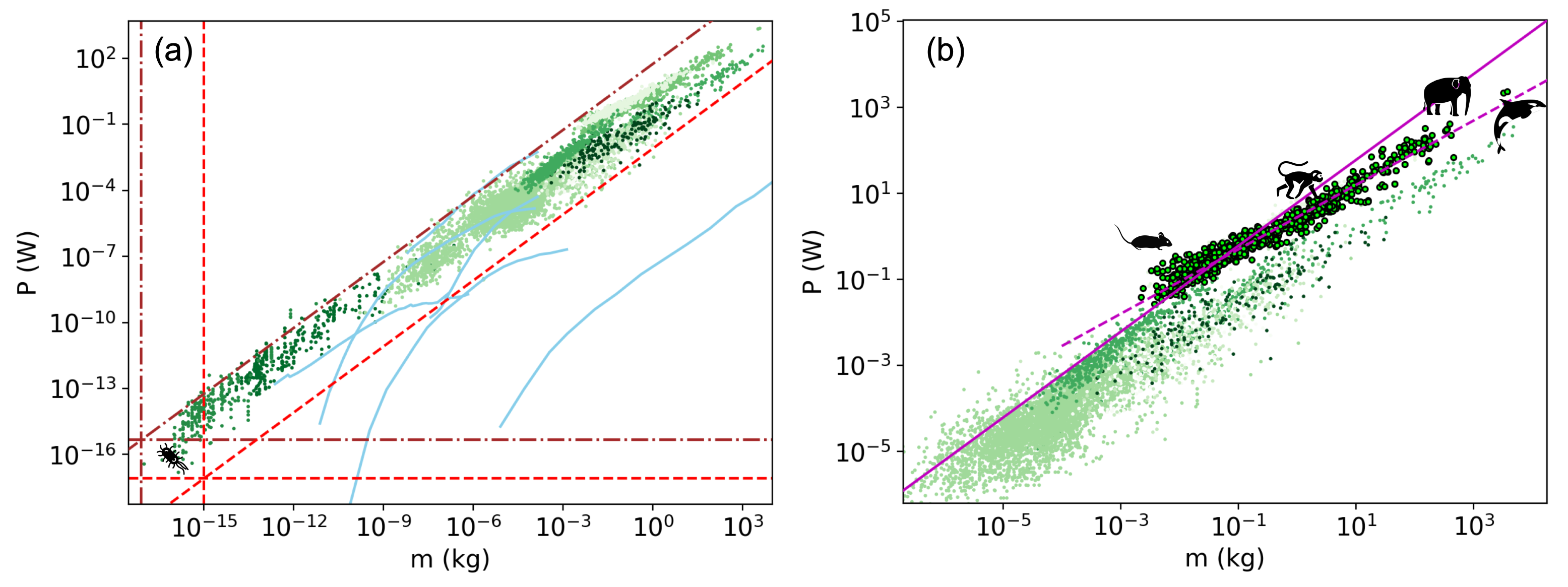}
\end{overpic}
\caption{Illustrations of the microscopic (a) and macroscopic (b) bounds of the metabolic rates of living organisms. (a) Two different choices of microscopic length $\ell_T$ lead to different values of the power density $\Pi$. Red dashed lines correspond to $\ell_T=1$~$\mu$m, whereas brown dotted-dashed lines correspond to $\ell_T=200$~nm. The blue lines reproduce some of the turbulence data from Fig.~\ref{fig1}. (b) Close-up on the metabolic rates of larger organisms. The highlighted symbols correspond to mammals, illustrating the crossover from $P\propto (\Pi/\rho) m$ (continuous line) to Kleiber's law, where $P\propto  \Pi^\frac{5}{2} \Psi^{-\frac{9}{4}} m^\frac{3}{4}$ (dashed line), assuming $\rho\simeq \rho_0$, $\Pi\simeq 6\Pi_0$ and $\Psi_0 =\rho_0 g_0\simeq 10^4$~kg.m$^{-2}$.s$^{-2}$. In (a) and (b) the different shades of green correspond to different taxonomic groups; see Hatton \textit{et al.} for details~\cite{Hatton2019}. }
\label{fig2}
\end{figure*}
In turbulence studies the microscopic length scale is variable, whereas for the metabolic rate its value can be set to the size of the smallest living organisms. Let us call $\ell_T$ this microscopic length, associated with a mass $m_T=\rho \ell_T^3$. Instead of using Kolmogorov's units to describe the microscopic scale, one may use `thermal units', based on the density $\rho$, the thermal energy $E=k_B T$, and the length $\ell_T$. Using dimensional analysis one can show that the associated time scale is: 
\begin{equation}
\tau_T \propto \Big(\frac{\rho \ell_T^5}{E} \Big)^\frac{1}{2}
\end{equation}
In this system of thermal units, one can obtain a power density: 
\begin{equation}
\Pi_T \propto m_T \ell_T^{-1} \tau_T^{-3} \propto \Big( \frac{E^3}{\rho \ell_T^{11}}\Big)^\frac{1}{2} 
\label{thermalPi}
\end{equation}
If we assume the reference density $\rho\simeq \rho_0$, room temperature such that $E\simeq 2~10^{-21}$~J, and a microscopic size $\ell_T\simeq 400$~nm, the thermal power density is $\Pi_T \simeq 10^3$~W/m$^3$, which is similar to the average $\Pi_0$ for all living organisms. As is apparent in Eq.~\ref{thermalPi}, the microscopic power density depends most sensitively on $\ell_T$. Changes of temperature or density have comparatively small effects. In Fig.~\ref{fig2}a, we give the values of power density predicted by the thermal units for $\rho\simeq \rho_0$ and room temperature, for $\ell_T\simeq 200$~nm or $\ell_T\simeq 1$~$\mu$m, which respectively give $\Pi_T \simeq 5.7~10^4$~W/m$^3$ and $\Pi_T \simeq 8$~W/m$^3$, a range that encompasses most of the metabolic rates of organisms. One may wonder what sets the value of $\ell_T$. This question is more easily answered by looking first at the other end of the spectrum. 

\section*{Macroscopic scale and gravity}
At the microscopic scale, no clear departure from $P\sim m$ could be identified, so we used $\ell_T$ to establish an expression of the power density based on the thermal energy. At the macroscopic scale, the metabolic data actually show significant departure from $P\sim m$, in particular for mammals, as shown in Fig.~\ref{fig2}b. These data sets are better represented by Kleiber's law, where $P\sim m^\frac{3}{4}$. To understand this scaling, one may return to an old question: why giants don't exist? This question has stimulated the development of dimensional analysis since its inception, and was famously addressed by Galileo~\cite{Galilei1914}. 

Schematically, the size of large living organisms can be understood from a balance between the weight, which tends to compress the organism, and the internal `pressure' maintaining the shape of the organism. The weight per unit volume is $\Psi=\rho g_0$ ($\mathcal{M}.\mathcal{L}^{-2}.T^{-2}$), where $g_0\simeq 10~$m.s$^{-2}$ is the acceleration of gravity. The `pressure' $\Sigma$ is the elastic modulus of the organism ($\mathcal{M}.\mathcal{L}^{-1}.T^{-2}$). The size is then given as $\ell_\star \propto \Sigma/\Psi$. Taking $\rho\simeq \rho_0$, and a modulus $\Sigma\simeq 10^5$~Pa, intermediate between muscles and bones, one gets $\ell_\star\simeq 10$~m. Of course, slightly different densities or elastic moduli would change this value, as is the case for trees.

In the framework of metabolic laws, Galileo's arguments can be adapted by looking for a relationship between power on one side, and mass $m$, force density $\Psi$ and power density $\Pi$ on the other. Dimensional analysis would lead to: 
\begin{equation}
P\propto \Pi^\frac{5}{2} \Psi^{-\frac{9}{4}} m^\frac{3}{4}
\label{PiPsi}
\end{equation}
This metabolic law is an example of Kleiber's law, where $P\sim m^\frac{3}{4}$. Assuming $\Pi\simeq 6\Pi_0$ and $\Psi_0 =\rho_0 g_0\simeq 10^4$~kg.m$^{-2}$.s$^{-2}$, this metabolic scaling is shown in Fig.~\ref{fig2}b to fit well with the metabolic rates of mammals. Slightly different values of $\Pi$ can be used to fit different taxonomic groups. Note that in this regime, $\Pi\neq \epsilon \rho$.  

Dimensionally, the ratio between a power density and a force density is a speed, $c\propto \Pi/\Psi$. In the case of mammals, this speed is $c\propto 6\Pi_0/\rho_0 g_0 \simeq 1~$m/s. Using this definition, one may rewrite Eq.~\ref{PiPsi} as:  
\begin{equation}
P \propto \Pi^{\frac{1}{4}} c^{\frac{9}{4}} m^{\frac{3}{4}} \propto \frac{\Pi}{\rho} m \Big(\frac{m_G}{m}\Big)^\frac{1}{4}
\label{Kleiber}
\end{equation}
Such law can also be obtained directly from dimensional analysis, since it is the only metabolic scaling depending solely on a power density $\Pi$ as well as on a characteristic speed $c$. In the right-most equation, this metabolic law is expressed as a correction to the linear scaling, where the macroscopic mass $m_G$ is defined as the intersection between this Kleiber's law and the linear scaling, such that $m_G\propto \rho^4 c^9 / \Pi^3 \propto \rho \ell_G^3$, with $\ell_G\propto c^3/\epsilon$. 

In analogy with Galileo's approach, one can assume that the crossover to Kleiber's law can be expressed as a ratio of elasticity and force density, $\ell_G\propto \ell_\star \propto \Sigma/\Psi$. If such assumption is combined with the definition $\ell_G\propto \rho c^3/\Pi$, then the speed $c$ can be expressed as a `sound speed':
\begin{equation}
 \frac{\rho c^3}{\Pi} = \frac{\Sigma}{\Psi} \Rightarrow c=\Big(\frac{\Sigma}{\rho} \Big)^\frac{1}{2}
\end{equation}
For mammals, the value $c\simeq 1~$m/s inferred empirically would give an elastic modulus $\Sigma\simeq 1$~kPa, which correspond to soft tissues. Invoking such speed of mechanical waves was central to some theoretical approaches to Kleiber's law~\cite{West2005}.

Overall, whereas the microscopic scale was well characterized by thermal units, the macroscopic scale is better described by what we might call `Galileo's units', depending on $\Psi$, $\rho$ and $\Sigma$. In addition to $\ell_G$ and $m_G$, the following time scale complements the system of units:  
\begin{equation}
\tau_G \propto \frac{(\Sigma\rho)^\frac{1}{2}}{\Psi}
\end{equation}
With these units, the power density can be defined macroscopically as: 
\begin{equation}
\Pi  \propto m_G \ell_G^{-1} \tau_G^{-3} \propto \Psi \Big(\frac{\Sigma}{\rho}\Big)^\frac{1}{2}
\label{macropow}
\end{equation}
Assuming $\Psi\simeq \Psi_0$ and $\rho\simeq \rho_0$ gives $\Pi\simeq 3~10^2 \Sigma^\frac{1}{2}$. For instance, if $\Pi\simeq \Pi_0$, this gives $\Sigma_0\simeq 10$~Pa.  

From this macroscopic perspective, the power density actually depends on the size of the host planet of the organisms, since $\Psi\propto \rho g\propto \rho G M/R^2$, where $G$ is the gravitational constant, and $M$ and $R$ are the mass and radius of the Earth. At the scale of the Earth, one can define an elasticity $\Sigma_E \propto GM^2/R^4\simeq 10^{12}$~Pa. This formula is obtained by applying Galileo's argument to the Earth, $R\propto \Sigma_E/\Psi$, which is known as the `hydrostatic equilibrium' in astrophysics~\cite{Carroll2017}. One can also define a time scale $\tau_E\propto (G\rho)^{-\frac{1}{2}}$~\cite{Carroll2017}. This time scale is sometimes called the `free-fall time' and interpreted as the time an object would take to collapse under its own gravity. Note that this time scale depends on the density rather than on the absolute size of the object, for a bacteria or the whole Earth it is around one hour. With these gravitational units, the macroscopic formula for the metabolic rate given in Eq.~\ref{macropow} can be written from a simple geometric mean of the elasticities of the Earth and of the tissues of the organisms: 
\begin{equation}
\Pi \propto \frac{(\Sigma \Sigma_E)^\frac{1}{2}}{\tau_E}
\end{equation}

\section*{A bridge between the molecule and the planet}
The average linear relationship between power and mass found for living organisms is a testimony of the intensive nature of the power density over a large range of sizes. In the microscopic realm of bacteria we have seen that the power density can be expressed using thermal units. In the macroscopic realm we have seen that linearity is broken due to the more dominant presence of gravity. If the power density is indeed intensive, the macroscopic and microscopic expressions ought to be equivalent:   
\begin{align}
&\Pi\propto \Big( \frac{E^3}{\rho \ell_T^{11}}\Big)^\frac{1}{2} \propto \Psi \Big(\frac{\Sigma}{\rho}\Big)^\frac{1}{2}\\
&\Rightarrow \ell_T \propto (\ell_0^9 \ell_G^2)^\frac{1}{11}
\end{align}
By equating the microscopic and macroscopic expressions of the power density, one can obtain a formula for the size of the smallest living organisms $\ell_T$ as a weighted geometric mean between the macroscopic length of Galileo's units, and the following length: 
\begin{equation}
\ell_0 \propto \Big( \frac{E}{\Sigma}\Big)^\frac{1}{3}
\end{equation} 
The form of this length scale can be understood if one recalls that an elasticity $\Sigma$ has the dimensions of an energy density. The length $\ell_0$ is then the scale at which the `elastic energy' $\Sigma\ell_0^3$ is equal to the thermal energy. By definition this length is smaller than that of the smallest living organism, i.e. $\ell_0<\ell_T$. Assuming room temperature and $\Sigma\simeq \Sigma_0$, this gives $\ell_0\simeq 60$~nm. This nanometric scale is at the crossroad of many physical phenomena combining thermodynamics, elasticity and electrostatics~\cite{Phillips2006}. It may be connected to the size of the `terminal units' used in some models of metabolism~\cite{West2005}. 

\section*{The metabolic outlook on turbulence}
So far, we have used the framework of turbulence to gain new insight on the metabolic relationships underlying the vast diversity of living organisms. In this last section, we would like to suggest how the metabolic framework can be used to study turbulence beyond Kolmogorov's `inertial regime'. 

In fluid turbulence, Kolmogorov's scaling is usually called `inertial', to underline the fact that the viscosity does not appear explicitly in the parameters of the scaling, which are $\Pi$ and $\rho$. How can living organisms display Kolmogorov's scaling? Living organisms do not abide to the Navier-Stokes equations underlying Kolmogorov's framework~\cite{Landau2013}. Nevertheless, they display a similar `inertia-like' spectrum because this scaling is more general than its particular instance in hydrodynamics. This generality is manifested most evidently by translating the power spectrum into a metabolic law, where the inertial spectrum is revealed as the expression of the intensive-to-extensive relationship, $P/m \propto \Pi/\rho$. `Non-inertial turbulence' is then a generic term for types of turbulence where this intensive-to-extensive relationship is broken, i.e. where either the density or power density can substantially vary. In these cases, the power spectra $\mathcal{E}(k)$ may be better expressed from quantities beyond $\rho$ and $\Pi$, ones that would be sufficiently constant to be used as parameters. Since Kolmogorov's studies in the 1940s, the phenomenology of turbulence has been applied to a very vast array of systems, which would be impossible to review here. To cite just a few examples, `non-inertial' regimes of turbulence have been described for compressible fluids~\cite{Andreopoulos2000} and magnetic fluids~\cite{Soward2005,Beresnyak2019}, including the interstellar medium~\cite{Armstrong1995,Fraternale2019}, for viscoelastic fluids~\cite{Groisman2000,Fouxon2003} and active fluids~\cite{Wensink2012,Kokot2017,Alert2020}, for waves~\cite{Newell2011,Falcon2010} or in two-dimensions~\cite{Boffetta2012}. For some of these systems, an understanding of the scaling of power spectra is still lacking. 

To the best of our knowledge, no review of the different observed energy spectra exist in the literature. This state of affair is unfortunate since some of these regimes are direct consequences of dimensional analysis and could be derived systematically. For instance, Kolmogorov's scaling $\mathcal{E} \sim k^{-\frac{5}{3}}$ is the scaling obtained by considering the parameters $\Pi$ and $\rho$. We have seen here that Kleiber's law corresponds to $\Pi$ and $\Psi$. If expressed as a power spectrum, Kleiber's law becomes: 
\begin{equation}
\mathcal{E} \propto \epsilon^{\frac{1}{6}} c^{\frac{3}{2}} k^{-\frac{7}{6}}
\label{Kleiberspectrum}
\end{equation}
We do not know if this power spectrum has been observed in the context of turbulence. We hope that readers will be able to answer this question. This power spectrum depends on $\Pi$ and $\rho$ through $\epsilon$, but also on $\Psi$ with the speed $c\propto \Pi/\Psi$. If one attempts to derive this power spectrum directly, as $\mathcal{E} \propto \epsilon^\alpha c^\beta k^\gamma$, no unique solution can be found. However, if one starts with a metabolic law $P\propto m^\alpha \Psi^\beta \Pi^\gamma$, a unique solution is found, and it can then be translated to a spectrum using Eq.~\ref{translate}. The mass dimension provides the additional constraint. As long as the density is constant, it can reappear in the power spectrum even if it was not used in the metabolic law, just by virtue of the fact that $k\propto (\rho/m)^\frac{1}{3}$. Under this condition, we can outline a general scheme, where one would pick two mass-carrying quantities $Q_1$ and $Q_2$, then seek the unique associated metabolic law $P\propto m^\alpha Q_1^\beta Q_2^\gamma$, and translate it to a power spectrum. 

To give one last example of this metabolic approach to turbulence, one can consider the metabolic law based on a pressure/elasticity $\Sigma$ and a density $\rho$. Dimensional analysis would lead to $P=m^{\frac{2}{3}}  \Sigma^{\frac{3}{2}} \rho^{-\frac{7}{6}}$. Such $\frac{2}{3}$ scaling has been discussed extensively in the metabolic literature, based on arguments of surface-to-volume ratios~\citep{Glazier2010}. We here provide a possible scaling for the prefactor. If expressed as a power spectrum, this law is simply $\mathcal{E}\propto c^2 k^{-1}$, with $c=(\Sigma/\rho)^\frac{1}{2}$ the sound speed. 

\section*{Conclusion}
Kolmogorov's spectrum is often described as formalizing the idea of the `energy cascade'. This view of turbulence as an energy cascade was famously expressed by Richardson~\cite{Richardson1926}: ``Big whirls have little whirls that feed on their velocity, and little whirls have lesser whirls and so on to viscosity''. In this picture, the energy is understood as cascading from the large scale where it is fed, all the way down to the viscous scale where it is dissipated as heat. In fact, Kolmogorov's spectrum in itself does not provide any direction in which the flow of energy is happening, just that the power density is constant over a wide range. The direction of the flow can only be inferred by higher order metrics~\cite{Josserand2017}. For simple fluids like water or air, the energy does cascade down. However, other types of turbulence can also be associated with inverse cascades, where the energy flows toward large scales~\cite{Boffetta2012}. 

For living organisms, we have seen that the metabolic rate can be understood in analogy with turbulence. The power density is constant over a wide range of scales, just as in Kolmogorov's scaling. One may ask in which direction the energy is flowing? In a way, the energy does seem to be `fed' at large scale, quite literally as ingurgitated calories. The bits and pieces of food are chewed down and broken up into their constituent molecules. Conversely, the metabolic chemical reactions involving these molecular nutriments generate an upward cascade, where thermal energy is progressively converted into work by molecular motors, cells, tissues, muscles, all the way up to the hand bringing the food to the mouth. Throughout living organisms, energy seems to be able to flow both ways. 

Looking back at Fig.~\ref{fig1}b or Fig.~\ref{fig2}a, the difference between the passive turbulence of fluids and the metabolic law of organisms seems to reside in what we may call the `pit of viscosity'. For passive fluids, the rate of energy transfer falls down before reaching the thermal scale we defined with bacteria. As the mass decreases, the power cascades down the pit of viscosity. In contrast, living organisms seem to throw a bridge over that pit to connect the molecular scale associated with $\ell_0$ and the astronomical scale associated with $R$. Life displays a roughly constant metabolic power density because it provides the smooth transition between the characteristic sizes and powers of the molecules and of the whole Earth. We hope that this new found connection will generate future studies on the relations between power and mass for objects beyond life, in order to refine the broad scaling relations we drew here, and to understand how life avoids the pit of viscosity. 

\textit{Acknowledgments}: The ideas of this paper were fostered by a lecture from Thomas Lecuit at Coll\`ege de France, by stimulating discussions with Mathieu Hautefeuille and Vivek Sharma, and by the joyful atmosphere of the Ladoux-M\`{e}ge lab. Joseph d'Alessandro is thanked for critically reviewing the manuscript. Ernesto Horne is thanked for his endorsement. 

\end{document}